\documentclass[nologo,url,11pt,a4paper]{ETHpaper}

\usepackage{graphicx}                  
\usepackage{amsmath, amssymb, amsfonts}
\usepackage{dcolumn}
\usepackage[usenames,dvipsnames]{xcolor}

\begin{document}
\title{Social Resilience in Online Communities: \\
  The Autopsy of Friendster}

\titlealternative{Social Resilience in Online Communities: The Autopsy
  of Friendster\\
  Submitted on February 22, 2013. \url{http://www.sg.ethz.ch/research/publications/}}
\author{David Garcia, Pavlin Mavrodiev, Frank  Schweitzer}

\author{David Garcia, Pavlin Mavrodiev, Frank  Schweitzer}
\address{Chair of Systems Design, ETH  Zurich, Weinbergstrasse 56/58, 8092 Zurich, Switzerland \\
  \url{dgarcia@ethz.ch,pmavrodiev@ethz.ch,fschweitzer@ethz.ch}}

\www{\url{http://www.sg.ethz.ch}}

\maketitle
\centerline{February 22, 2013}

\begin{abstract}

  We empirically analyze five online communities: Friendster,
  Livejournal, Facebook, Orkut, Myspace, to identify causes for the
  decline of social networks.  We define social resilience as the
  ability of a community to withstand changes. We do not argue about
  the cause of such changes, but concentrate on their impact. Changes
  may cause users to leave, which may trigger further leaves of others
  who lost connection to their friends. This may lead to cascades of
  users leaving. A social network is said to be resilient if the size
  of such cascades can be limited. To quantify resilience, we use the
  $k$-core analysis, to identify subsets of the network in which all
  users have at least $k$ friends.  These connections generate
  benefits (b) for each user, which have to outweigh the costs (c) of
  being a member of the network. If this difference is not positive,
  users leave.  After all cascades, the remaining network is the
  $k$-core of the original network determined by the cost-to-benefit
  ($c/b$) ratio.  By analysing the cumulative distribution of k-cores
  we are able to calculate the number of users remaining in each
  community. This allows us to infer the impact of the $c/b$ ratio on
  the resilience of these online communities.  We find that the
  different online communities have different k-core
  distributions. Consequently, similar changes in the $c/b$ ratio have
  a different impact on the amount of active users.  As a case study,
  we focus on the evolution of Friendster. We identify time periods
  when new users entering the network observed an insufficient $c/b$
  ratio. This measure can be seen as a precursor of the later collapse
  of the community. Our analysis can be applied to estimate the impact
  of changes in the user interface, which may temporarily increase the
  $c/b$ ratio, thus posing a threat for the community to shrink, or
  even to collapse.
\end{abstract}

\section{Introduction}

Online Social Networks (OSN), such as \texttt{Facebook} or
\texttt{Friendster}, can quickly become popular, but can also suddenly
lose large amounts of users.  The appearance of competing OSN, with
different functionalities and designs, create unexpected shifts of
users that abandon one community for another \cite{Giles2010a}. While
the dynamics of growth in these online communities are an established
research subject \cite{Backstrom2006, Kairam2012}, there are still
many open questions regarding the decline of online communities, in
particular related to large OSN \cite{Wu2013}. What are the reasons
behind the decision of users to stop using an OSN?  What is the role
of the social network in keeping user engagement, or in the spreading
of user dissatisfaction?  Are there network structures that lead to
higher risks of massive user departures?  In this article, we assess
the question of the relation between the topology of the user network,
and the cascades of user departures that threaten the integrity of an
online community. We build on previous theoretical work on network
effects \cite{Bhawalkar2012}, providing the first empirical study of
this phenomenon across successful, failed, and declining OSN.

The most successful OSN attract millions of users, whose interactions
create emergent phenomena that cannot be reduced back to the behavior
of individual users.  The OSN is a communication medium that connects
a large amount of people, which would stay together only if their
interaction dynamics leads to the emergent entity that we call
\emph{the community}.  The OSN and its users form a socio-technical
system in which the persistence of the community depends on both the
social interaction between users, and the implementation and design of
the OSN.  In this context, the \emph{social resilience}
\cite{Adger2000} of an online community is defined as \emph{``The
  ability of the community to withstand external stresses and
  disturbances as a result of environmental changes''}. In particular,
the technological component of the OSN can change the environment of
the users, and create stress that threatens the cohesion of the
community. As an example, changes in the user interface 
pose a general risk for user engagement in OSN.


The fast pace of the Internet society has already led to the total
disappearance of some very large online communities.  The most
paradigmatic example is \texttt{Friendster}, one of the first and
largest OSN, which suffered a massive exodus of users towards
competing sites. This led to its closure in 2011, to reopen as an
online gaming without its profile data. As a reaction, the
\texttt{Internet Archive}
\footnote{http://archive.org/details/friendster-dataset-201107}
crawled as much information as possible, creating a timeless snapshot
of \texttt{Friendster} right before its closure.  If, on the other
hand, \texttt{Friendster} was still an alive and active community,
this data would have been kept private and never made accessible at
such scale. Before closure, users were warned and offered to delete
their data from the site, leaving all the remaining data from this
community as one of the largest publicly available datasets on social
behavior.

The decay of \texttt{Friendster} is commented in a comedy video of the
Onion News, in which a fictitious \emph{``Internet archaelogist''}
explains \texttt{Friendster} as an ancient civilization
\footnote{https://www.youtube.com/watch?v=7mFJdOsjJ0k}. While proposed
as a satire of the speed of Internet culture, this video illustrates
the opportunities that a failed OSN offers for research. The users of
such a community leave traces that allow us to investigate its
failure.  In this sense, we can name our work as \emph{Internet
  Archeology}, because we analyze non-written traces of a disappeared
society, aiming at understanding the way it worked and the reasons for
its demise.

In this paper, we provide a quantitative approach to the collective
departure of users from OSN. We start from a theoretical perspective
that, under the assumption of rational user behavior, allows us to
define a new metric for the relation between network topology and
massive user leaves.  We apply this metric to high quality datasets
from \texttt{Friendster} and \texttt{Livejournal}, comparing their
social resilience with partial datasets from \texttt{Facebook},
\texttt{Orkut}, and \texttt{Myspace}.  The research presented here is
based on publicly available datasets, allowing the independent
validation of our results, as well as the extension to further
analyses \cite{Huberman2012}.  In addition, we focus on the time
evolution of \texttt{Friendster}, tracking the changes in its social
resilience and investigating how it decayed to a complete collapse. We
finish by commenting on the limitations and extensions of our
approach, and outline possible future applications.

\section{Related work}

Recent research has focused on the question of growth and decay of
activity or interest-based social groups \cite{Kleinberg2013}. This
line of research analyzes social groups as subcommunities of a larger
community, tied together due to underlying common features of their
members.  Such approach can be equally applied to scientific
communities and online social networks
\cite{Backstrom2006,Zheleva2009}, revealing patterns of diffusion and
homophily that respectively spread group adoption, and increase
internal connectivity. In particular, the big datasets provided by
online communities allow the study of group creation and maintenance
\cite{Kairam2012}, as well as the patterns of their internal network
structure across communities \cite{Laine2011,Yang2012}. These results
lead to applied techniques to predict the fate of interest-based
groups, and to improve clustering analysis of social networks.  Our
work differs from these previous results in the scope of our analysis:
Instead of looking at small to medium sized groups within larger
communities, we look at the OSN as a whole. In our approach, users are
not connected to each other due to certain common interest or
affiliation, but through an online platform that maintains their
social links and serves as communication medium.

Another research topic close to our work is the analysis of individual
churn, defined as the decision of a user to stop using a service in
favor of a competitor. This topic has received significant attention
due to its business applications in the telephony sector
\cite{Dasgupta2008}, studying how the social environment of an
individual can influence the switching to a competing company. In
addition, further studies explored how individual users disconnect
from P2P networks\cite{Herrera2007}, and stop using massive
multiplayer online games
\cite{Kawale2009}.

Regarding social networks, the question of user departure and churn
has special relevance \cite{Karnstedt2010}.  As an emerging topic, a
recent study shows the relation between social interaction and user
departure in the online community \texttt{Yahoo answers}
\cite{Dror2012}.  Furthermore, the same question has been addressed in
a recent article \cite{Wu2013}, analyzing a mysterious online social
network of which nor the name, size, nor purpose is explained. While
these results are relevant for the question of user engagement, it is
difficult to consider them in further research if we do not have
information about the nature of the studied network. Social networks
can have very different roles in online communities, requiring a
differentiation between traditional social networking sites
\cite{Laine2011}, and online communities with a social network
component, but where social interaction is mediated through other
channels. The results of \cite{Wu2013} reveal that 65\% of the users
that have no friends still remain active after three months,
indicating that such social network is not precisely necessary for a
user use the site. As an example, a \texttt{Youtube} user does not
need to create and maintain social contacts to interact with other
users, which can be done through videos and comments independently of
the social network.

Our work complements the previous results on individual user
departures mentioned above, as we analyze the social resilience of the
online community at the collective level. We build on these
empirically validated microscopic rules of churn, to focus on cascades
of departures through large OSN. We analyze the macroscopic topology
of the social network and its role in the survival of the
community. This kind of macroscopic effects are relevant to study the
emergence of social conventions \cite{Kooti2012}, an dynamics of
politically aligned communities \cite{Conover2012, Garcia2012}, in
addition to the case of OSN we address here.

The particular problem of enhancing resilience by fixing nodes of a
social network has been proposed and theoretically analyzed
\cite{Bhawalkar2012}, aiming to prevent the \emph{unraveling} of a
social network. This implies that social resilience can be analyzed
through the k-core decomposition of the social network, as explained
in Section \ref{sec:resilience} . In addition, k-core centrality is
the current state-of-the art metric to find influential nodes in
general networks \cite{Kitsak2010}, and information spreading in
politically aligned communities \cite{Conover2012}. Regarding social
media in general, the k-core decomposition was applied for a global
network of instant messaging \cite{Leskovec2008}, as well as for the
Korean OSN \texttt{Cyworld} \cite{Ahn2007, Chun2008}, motivated by
user centrality analysis rather than social resilience. To our
knowledge, this article introduces the first empirical analysis of
social resilience, relating changes in user environment with cascades
of departing users, through analysis based on the the k-core
decomposition of different OSN.

\section{Social Resilience in OSN}
\subsection{Quantifying Social Resilience}
A characteristic property of any online social network is the presence of
influence among friends. In particular, individual
decisions regarding participating or leaving the network are, to a large
extent, determined by the number of one's friends and their own
engagement \cite{Backstrom2006}. Therefore, users leaving a community have negative indirect
effects on their friends \cite{Wu2013}. This may trigger the
latter to also leave, resulting in further cascades of departing
users which may ultimately endanger the whole community. Social
resilience acts to limit the spread of such cascades. 

One approach to quantify social resilience is by natural removal
of nodes based on some local property, for example degree
\cite{Leskovec2008}. By studying the network connectivity after
such removals, one can identify nodes with critical importance for
keeping the community connected. Importantly, by focusing on local
properties we can only quantify the direct effects that a node removal
has on the connectivity of the network.

\begin{figure}[h]
  \centering
  \includegraphics[width=0.6\textwidth]{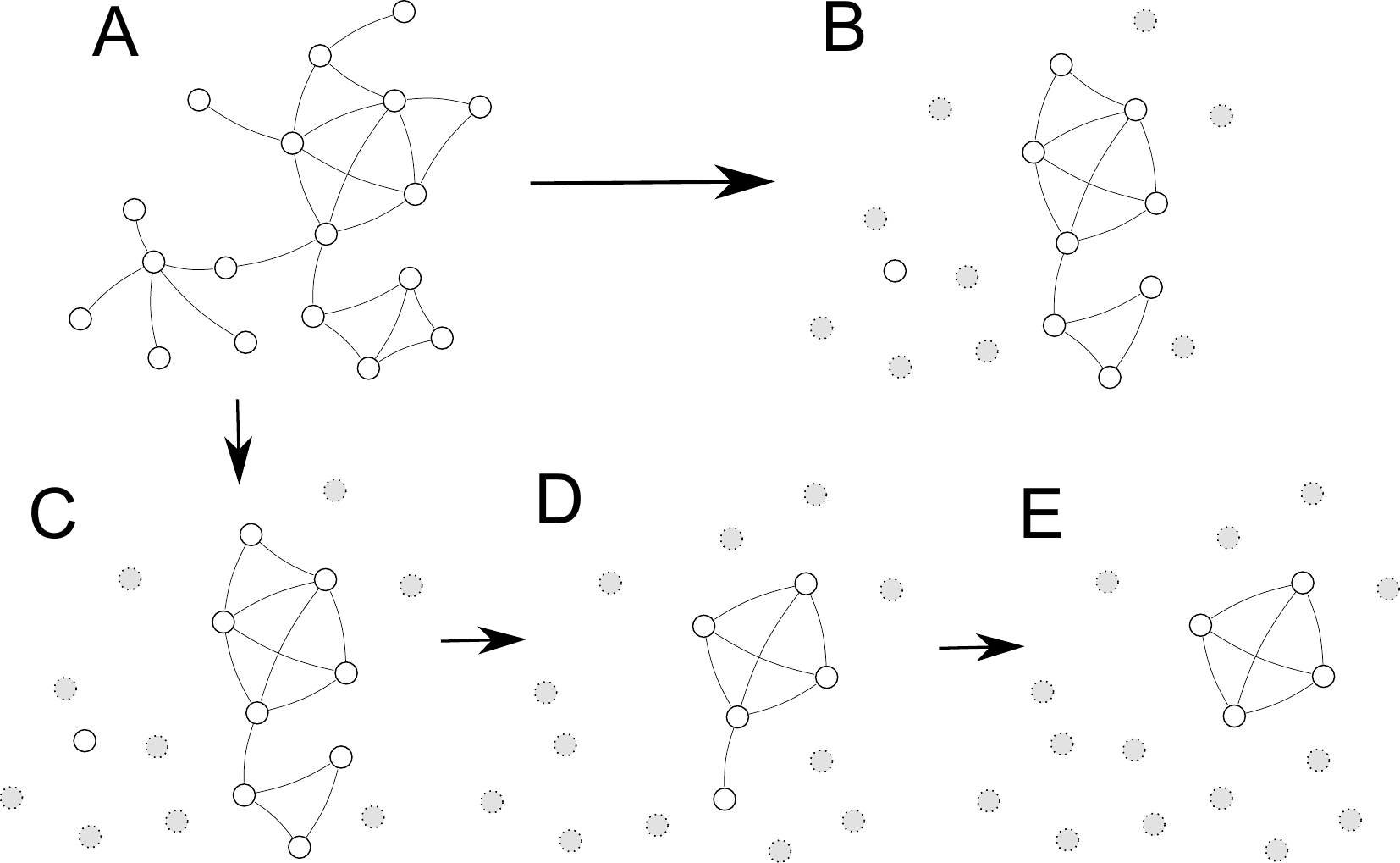}
  \caption{Effects of node removals on network connectivity as captured
    by degree only (A $\rightarrow$ B) and $k$-core decomposition (A
    $\rightarrow$ C $\rightarrow$ D $\rightarrow$ E)} 
  \label{fig:schema}
\end{figure}

In this paper, we propose an extension based on the $k$-core
decomposition \cite{Seidman1983}. A  $k$-core of a network is a sub-network in which
all nodes have a degree $\ge k$. The $k$-core decomposition 
is a procedure of finding all $k$-cores, $\forall k > 0$, by repeatedly
pruning nodes with degrees $k$. Therefore, it captures not only the
direct, but also the indirect impact of users leaving the network. As an
illustration 
consider Figure \ref{fig:schema}, which shows targeted removal of nodes
with degrees $<$ 3.
\label{sec:resilience}
On one hand, starting from the network in A and removing all nodes with degrees $<3$,
produces the network in B. The light-grey nodes in B have been removed
(and thus are disconnected), and the final network consists of the 9
white nodes. The transition A $\rightarrow$ B shows only the direct 
effects of users with $<$ 3 friends leaving.

On the other hand, starting again from A, and applying the $k$-core
procedure, will repeatedly remove nodes until only those with
degrees $\ge$ 3 remain. The first step, A $\rightarrow$ C, removes the
same light-grey nodes as before. Continuing, C $\rightarrow$ D, removes
those nodes that have been left with $<$ 3 neighbours in C, and
disconnects them as well. The final step, D $\rightarrow$ E, finishes the
process by disconnecting the last white node in D that was left with $<$ 3
friends. As a result, the final network is the fully connected network of the 
4 white nodes.  

Hence, supposing that users leave a community when they are left
with less than 3 friends, the $k$-core decomposition captures the full cascading
effect that departing users have on the network as a whole.

We proceed by formalizing social resilience based on a \emph{generalized}
$k$-core decomposition. To this end, we present a theoretical model in
which rational users decide simultaneously either to stay in the network or to leave it. These decisions are based on
maximizing a utility function that weighs the benefits of membership
against the associated costs. We show that the equilibrium network which maximizes the total
payoff in the community, corresponds to a
generalized k-core decomposition of the network. 
\subsection{Generalized k-core decomposition}

Following \cite{Harkins2013}, we extend the traditional $k$-core decomposition by
recognizing that the pruning criterion need not be limited to degree
only. Let us define a \emph{property} function $\mathcal{B}_{i}(H)$ that
given a sub-network $H \subseteq G$ associates a value, $n_{i} \in
\mathbb{R}$, to node $i$. A generalized $k$-core of a network $G$
is, then, defined as a sub-network $H \subseteq G$, such that
$\mathcal{B}_{i}(H) \ge k$, $\forall i \in H$ and $k \in \mathbb{Z}$. The
general form of $B_{i}$ allows us to model different pruning
mechanisms. For example, the traditional definition of the $k$-core can
be recovered in the following way -- for every node $i$ take its
immediate neighbourhood, $\mathcal{N}_{i}$, and fix
$B_{i}(H):=|\mathcal{N}_{i}|$, $\forall H \subseteq G$. Other authors
have also shown that considering weighted links in $\mathcal{B}_{i}$ can
more accurately reveal nodes with higher spreading potential in weighted
networks \cite{Garas2013}.

Note that by definition higher order cores are nested within lower order
cores. We use this to define that a node $i$ has \emph{coreness}
$k_{s}$ if it is contained in a core of order $k$, but not in a core of
order $k' > k$. 
\subsection{A rational model for OSN users}
\label{sec:model}
Here, we model the cost-benefit trade-off of OSN users in the
following way. Assume that users in a given network, $G$, incur a
constant integer cost, $c > 0$, for the effort they must spend to
remain engaged. Accordingly, they receive a benefit or payoff from
their friends in the network.  Let the benefit of player $i$ be the
property function $\mathcal{B}_{i}(H)$ with $i \in H$. Assume
non-increasing marginal benefits with respect to the size of $H$,
i.e. $\mathcal{B}_{i}''(H) \le 0$, otherwise costs are irrelevant as
any cost level could be trivially overcome by increasing the size of
$H$. This assumption is also supported by other empirical
investigations of large social networks which show that the
probability of a user to leave is concave with the number of friends
who left \cite{Backstrom2006,Wu2013}.

Players can choose one of two possible moves  -- \texttt{stay} or
\texttt{leave}. The utility of player $i$, is $U_{i}=0$, if he played
\texttt{leave} or $U_{i} = \mathcal{B}_{i}(H) - c$, for
\texttt{stay}. Finally, users are utility-maximizing, therefore they will
choose \texttt{stay} as long as $U_{i} > 0$.

It is easily seen that the equilibrium network, $G^{*}$, which maximizes
the total utility, $U(G) = \sum_{i}U_{i}$, is composed of users who
choose \texttt{stay} when $c<k^{i}_{s}$, and \texttt{leave} otherwise.
In other words, node $i$ should remain engaged in the network as long as
the cost, $c$, does not exceed its generalized coreness, $k_{s}$. In this
sense, $G^{*}$ corresponds to the
generalized $k$-core of $G$.  

To illustrate that $G^{*}$ is indeed an equilibrium network, we need
to show that no user has an incentive to unilaterally join it or leave
it. Consider a node, $j \in G^{*}$ who chooses \texttt{stay}. This
node would belong to a generalized $k$-core, $k^{j}_{s}$, and by
definition, $B_{j}(H) - k^{j}_{s} \ge 0$. Since, $j$ stayed in the
network, it must be that $c < k^{j}_{s}$, therefore $B_{j}(H)- c >
0$. So, $j$ will be forfeiting positive utility, had he decides to
leave. In the same manner, consider another node $l \notin G^{*}$ who
chooses \texttt{leave}, thus his coreness $k^{l}_{s} \le c$. All his
friends with the same coreness would have left the network, therefore
the only benefit that $l$ could obtain from staying would come from
his connections with nodes in higher cores. The benefit, $B_{l}$, from
such connections must not exceed $k^{l}_{s}$, otherwise $l$ would have
belonged to a higher core in the first place. Since $k^{l}_{s} \le c$
we have $B_{l} < c$. This implies that $l$ necessarily obtains
negative utility from staying, so he has no incentives to do
so. Moreover, $G^{*}$ is optimal, as we showed that any change from
the equilibrium actions of any user inevitably lowers his utility and
decreases the total utility in the network. We also argue that it is
reasonable to expect this equilibrium network to be reached in an
actual setting, since it maximizes the utility of all users
simultaneously, as well as the welfare of the network provider.

In the rest of the paper, we approximate $B_{i}$ as proportional to
the number of $i$'s direct friends, $N_{i}$, i.e. $B_{i} = bN_{i}$,
for some $b \in \mathbb{Z}$. Taking $k^{i}_{s}$ to be the coreness of
$i$, by definition it holds that $bN_{i} \ge k^{i}_{s}$. The maximum
cost, $c$, that $i$ would tolerate as a member of the community must
be strictly smaller than its coreness, hence $bN_{i} > c$ and $N_{i} >
c/b$. The last result implies that the minimum number of friends that
a node $i$ needs to remain engaged must be strictly larger than
$c/b$. Therefore, the coreness of a participating user $i$ must be at
least $c/b + 1$, i.e. $k^{i}_{s} \le K$, where capital $K=(c/b)+1$.

Based on the above discussion, we see that an user will remain in a
network with a high $c/b$ ratio if its coreness $k_{s}$ is high. This is
because, by definition $i$ is part of a 
connected network of nodes with large minimum degrees and hence large
benefits. 

In contrast, simply having a large degree does not imply that an user
will obtain large utility from staying. Note that a high-degree node
may nevertheless have low coreness. This means that $i$ would be part
of a sub-network in which all nodes have low minimum degrees. As a
result a lower $c/b$ ratio would suffice to start a cascade of users
departing, that can quickly leave $i$ with no friends and thus drive
it to leave too.

Finally, we define social resilience of a community as the size of the
$K$ core. In other words, this is the size of the network that remains
after all users with $k_{s} \le c/b$ have been forced out. This
definition allows us to quantify social resilience and reliably
compare it across communities even for unknown $c/b$ ratios, as shown
in Section \ref{sec:empirics}.

\section{Data on Online Social Networks}
\label{sec:data}
For our empirical study of social network resilience, we use datasets
from five different OSN. The choice of these datasets aims at spanning
a variety of success stories across OSN, including successful and
failed communities, as well as communities currently in decline.  The
size, data gathering methods, and references are summarized in Table
\ref{tab:data}, and outlined in the following.

\begin{table}[h]
\centering
\caption{Outline of OSN and datasets}
\label{tab:data}
\begin{tabular}{  |c|c|c|c|c|c|c|c|   } \hline
name&date&status&users&links&source\\ \hline

Livejournal & 1999 & \color{Green} successful & 5.2M & 28M &
\cite{Mislove2007} \\ \hline

Friendster & 2002 & \color{Red} failed  &  117M & 2580M &
\small{Internet Archive} \\ \hline

Myspace & 2003 & \color{Orange} in decline & 100K & 6.8M & \cite{Ahn2007} \\ \hline

Orkut & 2004 & \color{Orange} in decline & 3M & 223M  & \cite{Mislove2007} \\ \hline

Facebook & 2004 & \color{Green} successful &  3M & 23M & \cite{Gjoka2010} \\ \hline
\end{tabular}
\end{table}

\textbf{Friendster}\\
The most recent dataset we take into account is the one retrieved by
the Internet Archive, with the purpose of preserving
\texttt{Friendster}'s information before its discontinuation. This
dataset provides a high-quality snapshot of the large amount of user
information that was publicly available on the site, including friend
lists and interest-based groups \cite{Yang2012}. In this article, we
provide the first analysis of the social network topology of
\texttt{Friendster} as a whole.

Since some user profiles in \texttt{Friendster} were private, this
dataset does not include their connections.  However, these private
users would be listed as contacts in the list of their friends who
were not private. We symmetrized the \texttt{Friendster} dataset by
adding these additional links. Due to the large size of the
\texttt{Friendster} dataset, we symmetrized the data by using Hadoop,
which we distribute under a creative commons license
\footnote{web.sg.ethz.ch/users/dgarcia/Friendster-sim.tar.bz2}.

\textbf{Livejournal}\\
In \texttt{Livejournal}, users keep personal blogs and define
different types of friendship links. The information retrieval method
for the creation of this dataset combined user id sampling with
neighborhood exploration \cite{Mislove2007}, covering more than 95\%
of the whole community. We choose this \texttt{Livejournal} dataset
for its overall quality, as it provides a view of practically the
whole OSN.

Note that the desing of \texttt{Livejournal} as an OSN deviates from
the other four communities analyzed here. First, \texttt{Livejournal}
is a blog community, in which the social network functionality plays a
secondary role. Second, \texttt{Livejournal} social links are
directed, in the sense that one user can be friend of another without
being friended back. In our analysis, we only take include reciprocal
links, referring to previous research on its k-core decomposition
\cite{Kitsak2010}. By including this dataset, we aim at comparing how
different interaction mechanisms and platform designs influence social
resilience.

\textbf{Orkut}\\
Among declining social networking sites, we include a partial dataset
on \texttt{Orkut} \cite{Mislove2007}, which was estimated to cover
11.3\% of the whole community. Far from the quality of the two
previous datasets, we include \texttt{Orkut} in our analysis due to
its platform design, as this dataset includes users that did not have
a limit on their amount of friends. Furthermore, \texttt{Orkut} has a
story of local success in Brazil, losing popularity against other
sites at the time of writing of this article. 

\textbf{MySpace}\\
One of the most famous OSN in decline is \texttt{Myspace}, which was
the leading OSN before \texttt{Facebook}'s success \cite{Giles2010a}.
We include a relatively small dataset of $100000$ users of MySpace
\cite{Ahn2007}, which was aimed to sample its degree
distribution. This dataset was crawled through a Breadth-First Search
method, providing a partial and possibly biased dataset of
\texttt{Myspace}. We include this dataset as an exercise to study the
influence of sampling biases in the analysis of social resilience.

\textbf{Facebook}\\
We want to complete the spectrum of success of OSN, from the collapse
of \texttt{Friendster} to the big success of \texttt{Facebook}. The
last dataset we include is a special crawl which aims at an unbiased,
yet partial dataset as close as possible to the whole community
\cite{Gjoka2010}. This dataset was retrieved with a special technique
based on random walks, keeping unvaried some network statistics,
including \texttt{Facebook}'s degree ditribution.

\section{Not power-law degree distributions}

The first step in our analysis explores the degree distributions of
each OSN. The reason to do so is the epidemic properties of complex
networks. Under the assumptions of epidemic models, networks with
power-law degree distributions do not have an epidemic threshold
\cite{Pastor-Satorras2002}, i.e. a ``sickness'' would survive within
the network for an unbound amount of time and eventually infect most
of the nodes. Such sickness could be a meme or a social norm, but
could also be the decision of leaving the community. Therefore, we
need to assess the posibility of a lower-law degree distribution, as
it would pose an alternative explanation for the masive cascades of
user departures.

Numerous previous works have reported power-law degree distributions
in social networks \cite{Ahn2007, Chun2008, Leskovec2008,
  Mislove2007}. Nevertheless, most of these works rely on goodness of
fit statistics, and do not provide a clear test of the power-law
hypothesis. It states that the degree distribution follows the
following equation $p(d) = \frac{\alpha-1}{\text{deg}_{\min}} \left (
  \frac{d}{\text{deg}_{\min}} \right ) ^{-\alpha} $ for $d \geq
\text{deg}_{\min}$. This is usually described as $p(d) \propto
d^{-\alpha}$, and often argued as valid if metrics such as $R^2$, or
$F$ are high enough. While a high goodness of fit could be sufficient
for some practical applications, the empirical test of the power-law
hypothesis can only be tested, and eventually rejected, through the
result of a statistical test, assuming a reasonable confidence level.

\begin{figure}[ht]
  \centering
  \includegraphics[width=0.8\textwidth]{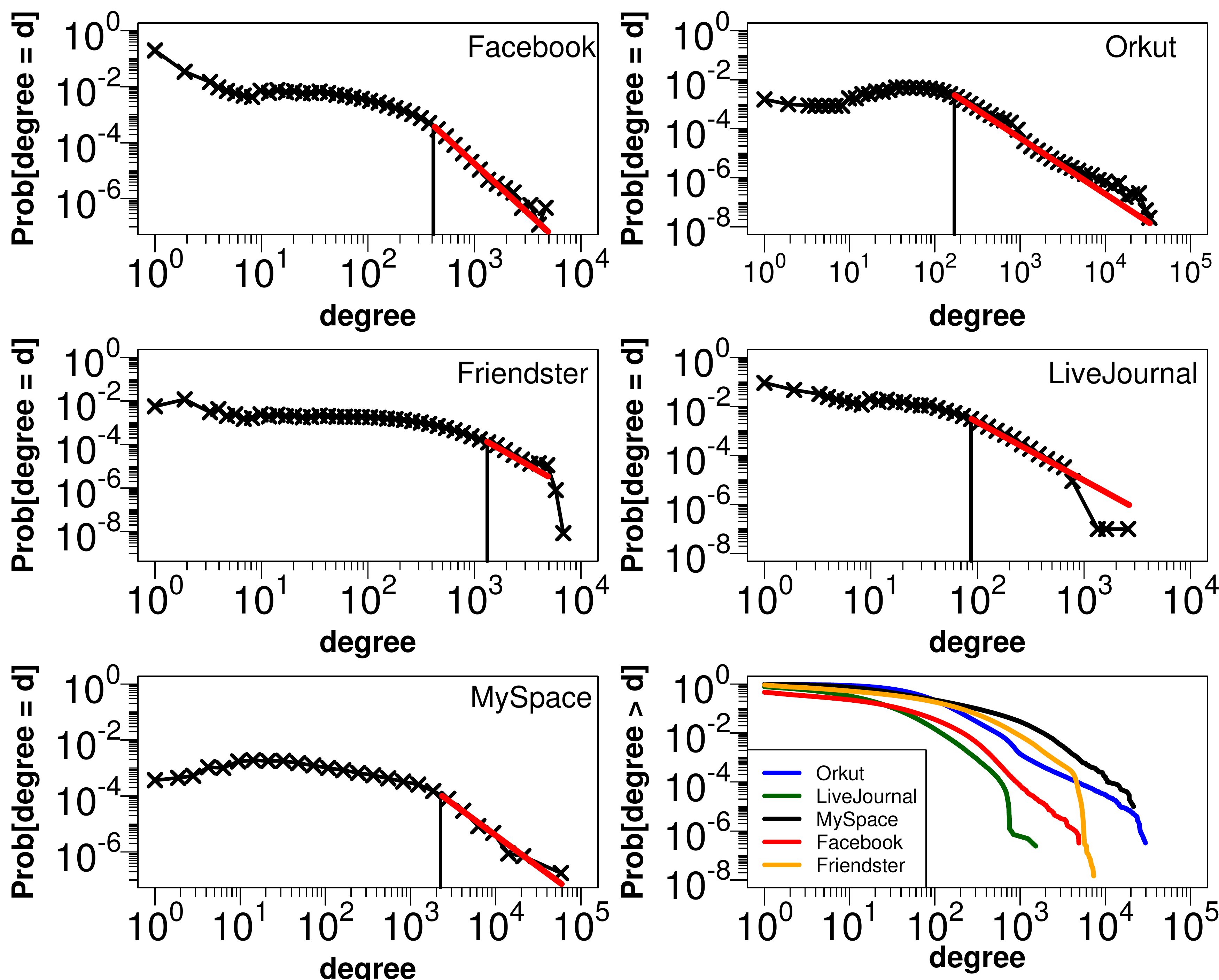}
  \caption{Complementary cumulative density function and probability
    density functions of node degree in the five considered
    communities. Red lines show the ML power-law fits from
    $\widehat{\text{deg}}_{\min}$}
  \label{fig:fancy-plot}
\end{figure}

We followed the state-of-the-art methodology to test power laws
\cite{Clauset2009}, which roughly involves the following steps. First,
we created Maximum Likelihood (ML) estimators $\hat{\alpha}$ and
$\widehat{\text{deg}}_{\text{min}}$ for $p(d)$. Second, we tested the
empirical data above $\widehat{\text{deg}}_{\text{min}}$ against the
power law hypothesis and we recorded the corresponding KS-statistics
($D$). Third, we repeated the KS test for 100 synthetic datasets that
follow the fitted power law above
$\widehat{\text{deg}}_{\text{min}}$. The p-value is then the fraction
of the synthetic $D$ values that are larger than the empirical
one. 
Thus, for each degree distribution, we have the ML estimates
$\widehat{\text{deg}}_{\min}$ and $\hat{\alpha}$, which define the
best case in terms of the KS test, with an associated $D$ value,
and the p-value.

Ultimately, a power law hypothesis cannot be rejected if (i) the
p-value of the KS-test is above a chosen significance level
\cite{Clauset2009}, and (ii) there is a sufficiently large amount of
datapoints from $\text{deg}_{\min}$ to $\text{deg}_{\max}$
\cite{Stumpf2012}. We found that the degree distributions of
\texttt{Facebook}, \texttt{Friendster}, \texttt{Orkut} and
\texttt{Livejournal} have p-values well below any reasonable
significance threshold, showing an extremely reliable empirical
support to reject the power-law hypothesis (Table \ref{table:pfits}).

For the case of \texttt{Myspace}, a KS test gives a p-value of $0.22$,
which can be considered high enough to not reject the power-law
hypothesis \cite{Clauset2009}. Therefore \texttt{Myspace} satisfies
the first criterion, but when looking at the range of values from
$\text{deg}_{\min}$ to $\text{deg}_{\max}$ (roughly one order of
magnitude), and the low amount of datapoints included, this KS-test
composes a merely anecdotal evidence of the extreme tail of Myspace.
If accepted, the power-law distribution would explain just 0.623\% of
the \texttt{Myspace} dataset.  In addition, the unsupervised BFS
crawling method used for this dataset has been shown to have a bias
that creates artificial power-law tails \cite{Achlioptas2005}. This
leads to the conclusion that, while we cannot fully reject the
power-law hypothesis, we can safely state that the dataset does not
support the hypothesis otherwise.  Figure \ref{fig:fancy-plot} shows
the degree distributions and their CCDF. For each OSN, we show how the
typical log-log plot of the PDF is misleading, as a simple eye
inspection would suggest power-law distributions, but a robust
statistical analysis disproves this possibility.

\begin{table}[h]
\centering
\caption{Power law fits of the analyzed datasets.}
\begin{tabular}{c|ccc|cc}
\label{table:pfits}
dataset  &
$\widehat{\text{deg}}_{\min}$ & $\hat{\alpha}$ & $n_{\rm tail}$ &$D$&$p$\\
\hline
Friendster   & 1311 & 3.6 & $2.9 \times 10^{5}$ &4.59& $<10^{-15}$\\ 
LiveJournal  & 88 & 3.3 & 81141  &0.02& $<10^{-15}$\\ 
Facebook     & 423 & 4.6 & 4918 &0.14& $<10^{-15}$\\ 
Orkut        & 171 & 3 & $2.8 \times 10^{5}$ &0.02& $<10^{-15}$\\ 
MySpace      & 2350 & 3.6 & 623 &0.03& 0.22\\ 
\end{tabular}
\end{table}

\section{Empirics of OSN Resilience}
\label{sec:empirics}
\subsection{K-core decomposition}

\begin{figure}[h]
  \centering
  \includegraphics[width=0.75\textwidth]{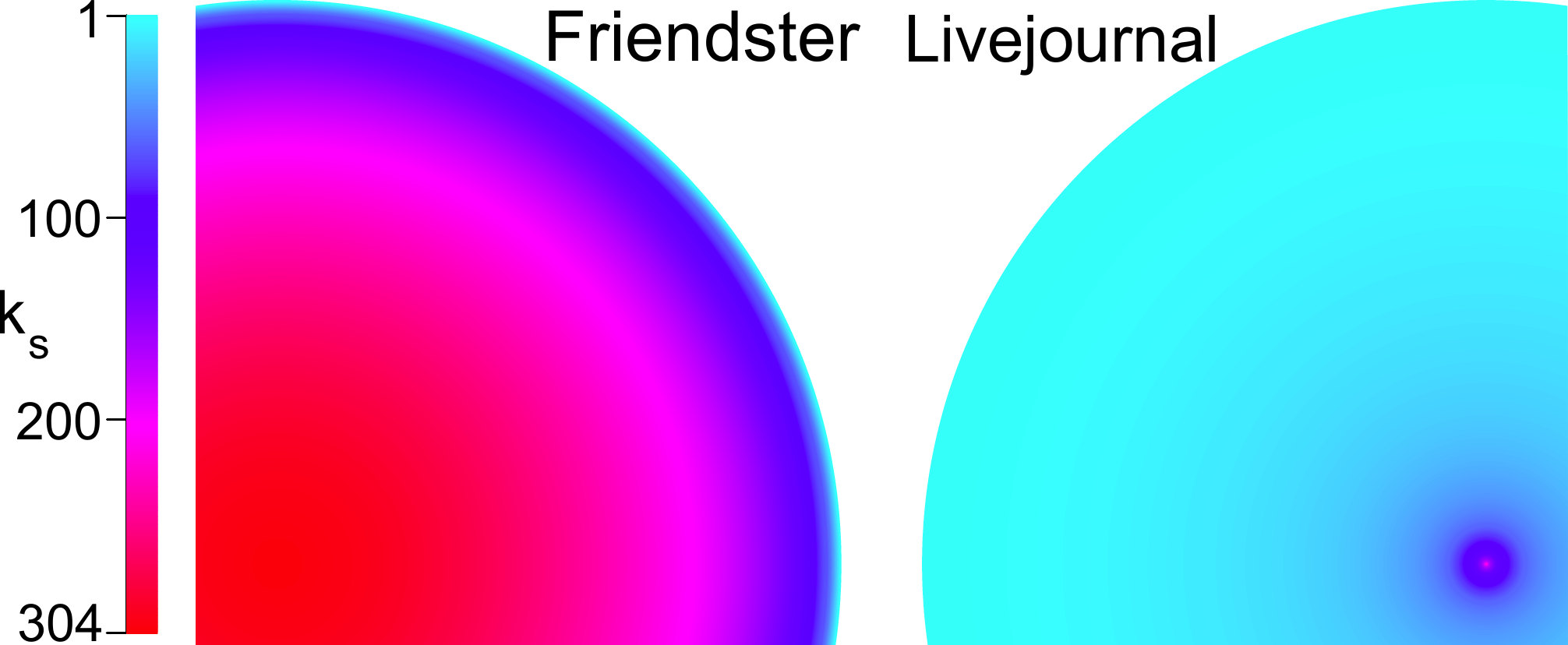}
  \caption{Overview of the k-core decomposition for Friendster and
    Livejournal.  Layers are colored according to $k_s$, with areas
    proportional to the amount of nodes with such $k_s$. }
  \label{fig:coreCircles}
\end{figure}

We computed the k-core decomposition for each of the OSN datasets we
introduced in Section \ref{sec:data}.  Among those datasets,
\texttt{Friendster} and \texttt{Livejournal} cover the vast majority
of their respective communities. Figure \ref{fig:coreCircles} shows a
schematic representation of the k-cores of \texttt{Friendster} and
\texttt{Livejournal}. Each layer of the circles corresponds to the
nodes with coreness $k_s$, with an area proportional to the amount of
nodes with that coreness value. The color of each layer ranges from
light blue for $k_s=1$, to red for $k_s=304$. The distribution of
colors reveals a qualitative difference between both communities:
\texttt{Friendster} has many more nodes of high coreness than
\texttt{Livejournal}, which has a similar color range but a much
larger fringe, i.e. the set of nodes with low $k_s$. This difference
indicates that, to keep together as a community, \texttt{Livejournal}
needs to have a much lower $c/b$ than \texttt{Friendster}. This
scenario is rather realistic, as \texttt{Livejournal} is a blog
community in which users create large amounts of original
content. This leads to high benefits per social link as long as users
have similar interests, which seems to be the key of
\texttt{Livejournal}'s relative success.

\begin{figure}[h]
  \centering
\centerline{  \includegraphics[width=0.45\textwidth]{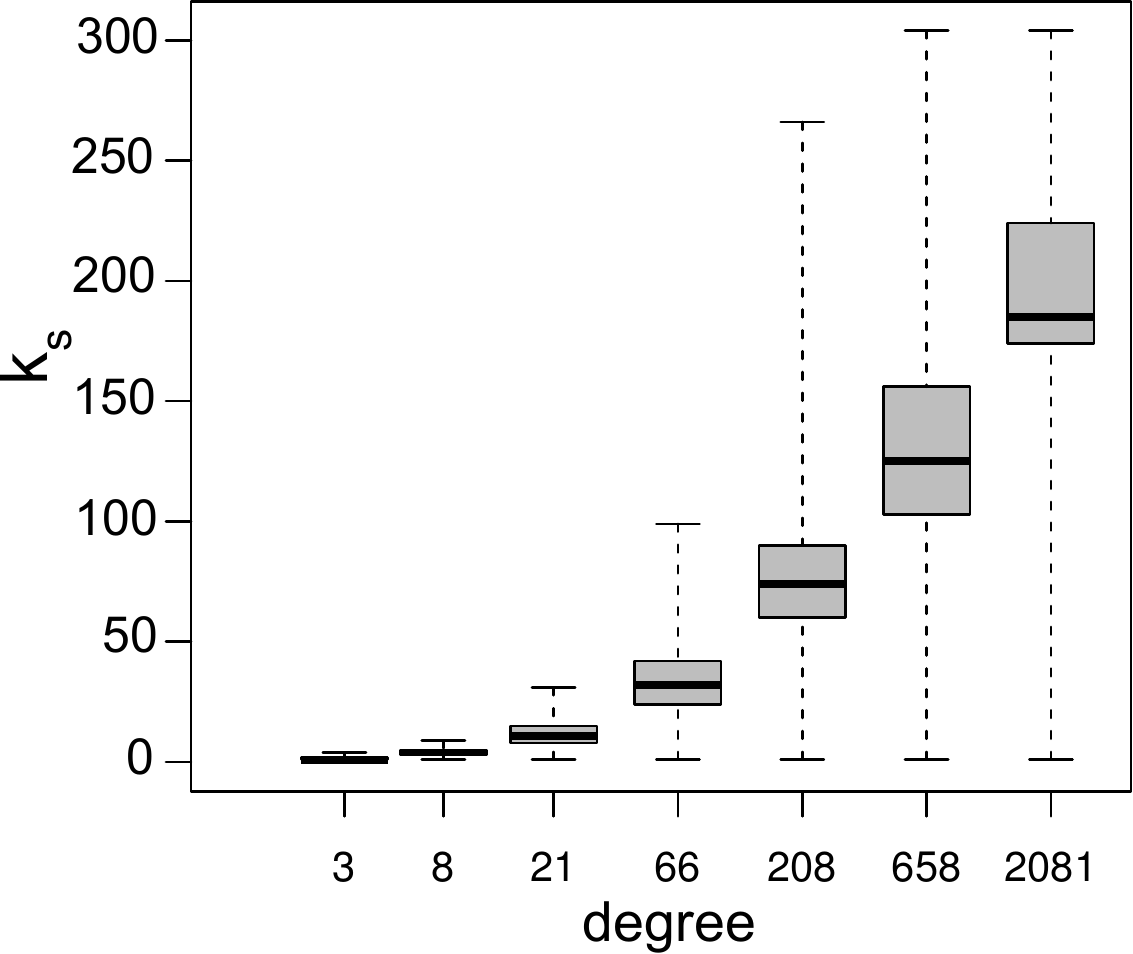}
  \hfill \includegraphics[width=0.51\textwidth]{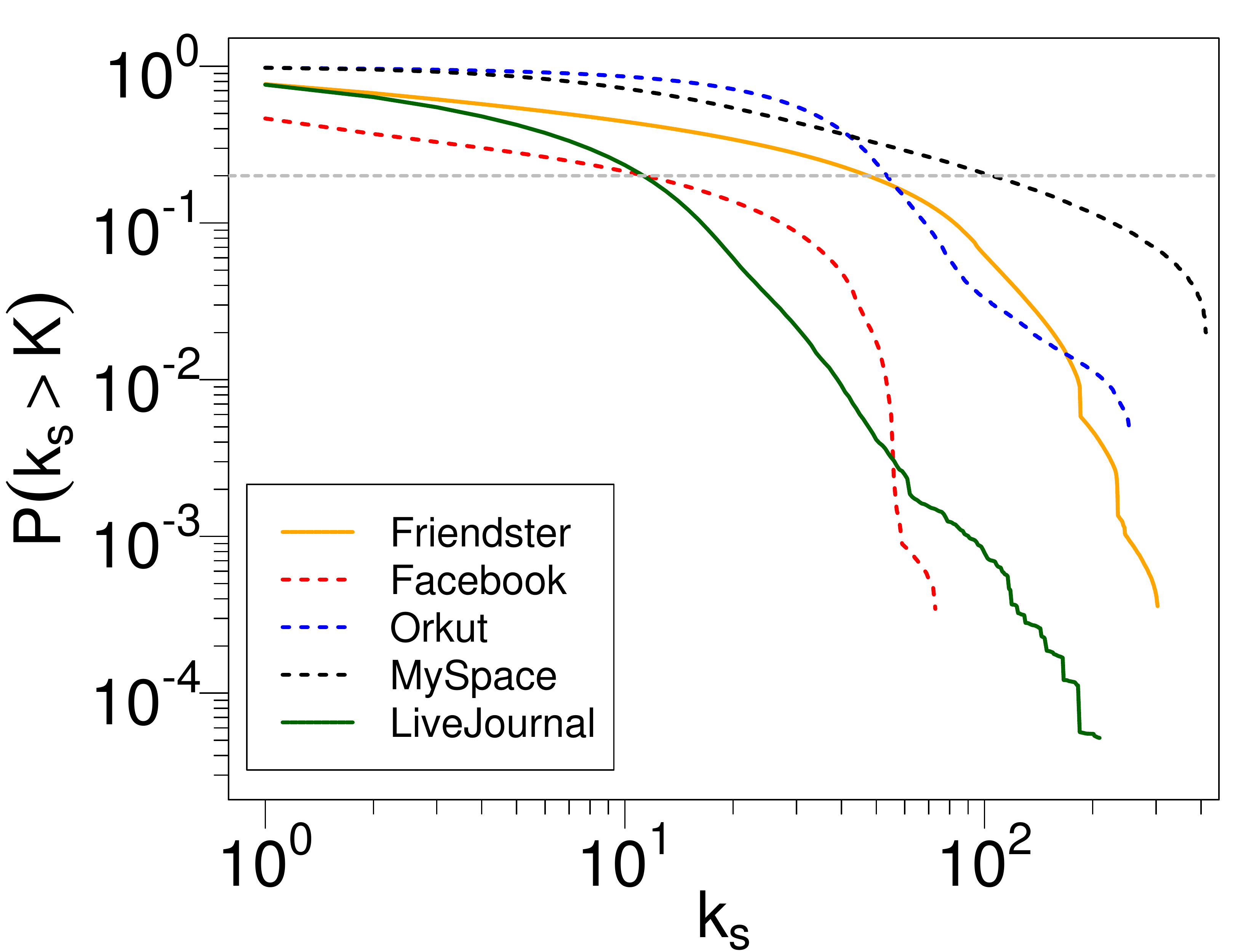}}

  \caption{Left: boxplot of k-shell indices by degree for
    \texttt{Friendster}. Dark lines represent the mean, and dashed bars show extreme
    values. Boxes are arranged in the x-axis according to the middle
    value of their bin. Right: CCDF of $k_s$.  The horizontal dashed
    line shows the cut at 0.2. }
  \label{fig:coreDeg}
\end{figure}

Our theoretical argumentation, presented in Section \ref{sec:model},
indicates that node coreness is a more reasonable estimator for
resilience than node degree. A degree of at least $k_s$ is a necessary
condition for a coreness of $k_s$, but a high degree does not
necessarily mean a high coreness.  Taking \texttt{Friendster} an
example, Figure \ref{fig:coreDeg} shows the boxplot for the
distribution of $k_s$ versus node degree, indicating the spread of
$k_s$ for nodes of similar degree. The empirical data shows that a
high degree does not necessarily mean a high $k_s$, even finding nodes
with very low $k_s$ and very high degree. Nevertheless, it is clear
that $k_s$ is likely to increase with degree, but mapping degree to
coreness would misestimate the resilience of the community as a whole.
By measuring coreness, whe can detect that some nodes belong to the
fringe despite their high degree, as the coreness integrates global
information about the centrality of the node.

\subsection{Resilience comparison}
Extending the above observations, we computed the k-core decomposition
of the three additional OSN, aiming at comparing their relation
between their environment, measured through $c/b$, and the amount of
users expected to be active under such conditions.

We focus our analysis on the Complementary Cumulative Density Function
(CCDF) of each network, defined as $P(k_s>K)$. As shown in Section
\ref{sec:model}, the cost-benefit-ratio $c/b$ corresponds to a value $K$ that
determines the nodes that leave the network, which are those $k_s$
coreness below $K$. Under this conditions, the CCDF of $k_s$ measures
the amount of nodes that will remain in the network under a given
$c/b$, allowing us to compare how each OSN would withstand the same
values of cost and benefit.

The right panel of Figure \ref{fig:coreDeg} shows the log-log CCDF
of the five OSN. The first two communities to compare are
\texttt{Livejournal} and \texttt{Friendster}, as the datasets on these
two are the most reliable. First, the CCDF of \texttt{Friendster} is
always above the CCDF of \texttt{Livejournal}. This is consistent with
the structure shown in Figure \ref{fig:coreCircles}, where it can be
appreciated that \texttt{Livejournal} has many more nodes in the
fringe than \texttt{Friendster}. Second, both CCDF reach comparable
maximum values, regardless of the fact that \texttt{Friendster} was 20
times larger than \texttt{Livejournal}.  Such skewness in the coreness
of \texttt{Livejournal} can be interpreted as a result of a higher
competition for attention, as expected from a blog community in
comarison with a pure social networking site, like \texttt{Friendster}
was.

Focusing on the tails of the distributions, we can compare the
patterns of resilience for environments with high $K$. The comparison
between the resilience of these communities is heavily dependent of
the value of $K$, as for example, \texttt{Livejournal} is less
resilient than \texttt{Facebook} for values of $K$ between 10 and 50,
but more resilient below and above such interval. A similar case can
be seen between \texttt{Friendster} and \texttt{Orkut}, as their CCDFS
cross at 60 and 200. Thus, \texttt{Friendster} would be more resilient
than \texttt{Orkut} if $K$ lies in that interval, while \texttt{Orkut}
would have a larger fraction of active nodes if $K<60$ or $K>200$.

It is important to note that these comparisons are made between the
reliable datasets of \texttt{Friendster} and \texttt{Livejournal},
compared with partial datasets from the other communities. While our
conclusions on the first two OSN can be seen as global findings on the
community as a whole, the rest are limited to the size of the
datasates available. A particularily clear example of the effect of
the crawling bias is the distribution of coreness for
\texttt{Myspace}, which shows an extreme resilience in comparison to
all the other datasets, with the exception of \texttt{Orkut} for
$K<50$. As commented in Section \ref{sec:data} , the method used for
\texttt{Myspace} was very biased towards nodes of high degree, leaving
an unrealistic picture of the resilience of the whole
community. Additionally, the method used for \texttt{Facebook} seems
to have delivered a degree distribution close to a random sample of
\texttt{Facebook} users, but its restarting of random walkers leaves
tendrils of nodes that accumulate on the 1-core. Hence the low
starting value of the CCDF of \texttt{Facebook} could be an artifact
of this crawling method.

Regardless of any crawling bias, we found that these networks have
maximum coreness numbers much higher than previous results. The
maximum $k_s$ found for the network of instant messaging was limited
to 68 \cite{Leskovec2008}, and close to 100 for the OSN
\texttt{Cyworld} \cite{Chun2008}.  \texttt{Livejournal} has a maximum
$k_s$ of 213, \texttt{Friendster} of 304, \texttt{Orkut} of 253, and
\texttt{Myspace} as a very deep core of $k_s=414$. The exception lies
in the \texttt{Facebook} dataset, where we find a maximum $k_s$ of
74. This evidence shows that OSN can have much tigther cores than the
ones found in previous research, revealing that they contain small
communities with very high resilience.

As a final comparison, we focus on the values of $K$ for the
catastrophic case of the networks losing 80\% of their nodes,
i.e. where the CCDF has a value of 0.2. The data shows that both
\texttt{Facebook} and \texttt{Livejournal} would lose 80\% of their
users under a value of $K$ close to $10$.  For the case of the
unsuccessful communities of \texttt{Orkut} and \texttt{Friendster}, it
requires a much worse environment, with values of $K$ above $60$.
This way, the emprical data supports the idea that, under the same
environmental conditions, both successful communities is less
resilient than the three unsuccessful ones. This means that the
topology of their social network is not enough to explain their
collapse, but indicates that bad decisions in design and interface
changes can spread through the network and drive many users away.

\section{The Time Evolution of Friendster}

In this section, we describe a \emph{post hoc} case study of the way
how \texttt{Friendster} rised and collapsed, using the avaliable
timing information in the dataset.

\subsection{Social growth mechanism}
The \texttt{Friendster} dataset does not provide the date of creation
of user accounts or social links, but it includes a user id that
increased sequentially since the creation of the site.  We analyzed
the time series of \texttt{Friendster} in an event time scale, where
each timestamp corresponds to the id of each user. We measured the
time distance of an edge $e$, which connects users $u_1$ and $u_2$, as
the difference between the ids of these users $d(e(u_1,u_2)) =
|id(u_1) - id(u_2)|$. In the following, we how early users connected
to later users, making the network grow.

We divided the network in time slices of a width of 10 million user
ids, with a last smaller slice of 7 million ids. Each of these 12
slices contains a set of nodes that have connections i) to nodes that
joined the community before, ii) to nodes that joined the network
afterwards, and iii) internally within the slice. This way, for the
slice of time period $t$ we can calculate its internal average degree
$2|E_{in}(t)|/ |N(t)|$, where $E_{in}(t)$ is the set of edges between
nodes in the slice $t$, noted as $N(t)$.

As an extension, we define $E_p(t)$ and $E_f(t)$ as the sets of edges
towards nodes that joined the community before $t$ (past nodes), and
nodes that joined after $t$ (future nodes). We measured the time range
of connections to the past $P(t)$ as the mean distance of the edges in
$E_p(t)$, and the rage of connections to the future $F(t)$ as the mean
distance of their future counterpart $E_f(t)$.  By definition, the
amount of past nodes for the first slice is $0$, equally to the amount
of future nodes for the last slice. If the process of edge creation
was purely random, the network would resemble an Erd\"os-Renyi graph
with an arbitrary sequence of node ids. In such network, $P(t)$ would
steadily increase with each slice, having an expected value of $|N|/2$
for the last one, where $|N|$ is the size of the network. Similarly,
$F(t)$ would decrease from $|N|/2$ at the first slice, converging at
$0$ in the last one.

\begin{figure}[h]
  \centering
    \includegraphics[width=0.65\textwidth]{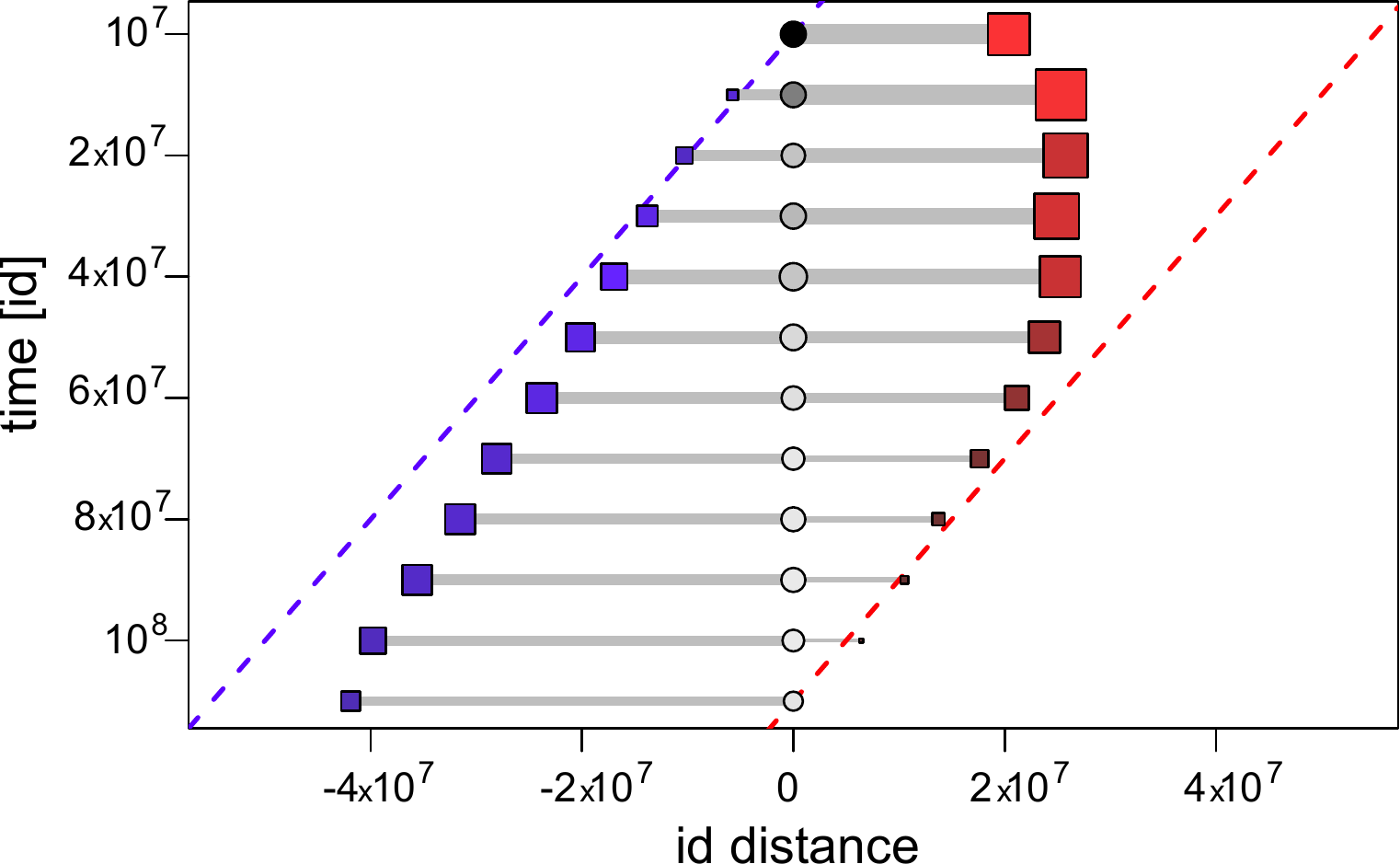}
    \caption{Schema of connectivity of Friendster users across
      time. Each circle represents a slice of the network of width of
      10 million user ids. Blue squares represent past users and red
      squares represent future users, with a distance from their slice
      according to $P(t)$ and $F(t)$ respectively. The dashed lines
      show the expectation of these two metrics for a random graph.}
  \label{fig:timeplot}
\end{figure}

The time evolution of the range of connections to past and future is
shown in Figure \ref{fig:timeplot}.  Each circle represents a slice of
the network, with growing $t$ from top to bottom. Their vertical
alignment represents the present with respect to the slice, and each
circle is connected to a blue square on the left that represents past
nodes, and a red square on the right that represents future nodes.
Balls have a size proportional to $|N(t)|$, which keep approximately
constant throughout time. The darkness of each circle is proportional
to its internal connectivity $|E_{in}(t)| $, and the width of the
connections from circles to past and future squares are proportional
to $|E_p(t)|$ and $|E_f(t)|$ respectively.  Internal connectivities
decrease through time, as early slices had significantly higher
$|E_{in}(t)|$. This indicates that the initial root of users of
\texttt{Friendster} was much more tightly connected among themselves
than towards other nodes, creating a denser subcommunity of old
users. A possible explanation for this pattern is that
\texttt{Friendster} started as an OSN for dating, and its design was
later shifted towards generalized networking as it became popular.

The squares of Figure \ref{fig:timeplot} are positioned according to
the mean past $P(t)$ and future $F(t)$ distances of each slice. As a
comparison with random network construction, dashed lines show their
expected values as explained above. For early slices, the mean future
distance is significantly lower than its random counterpart, revealing
a pattern of time cohesion that limits the range of future
connections. This shows a decay in the diffussion process through the
offline social network, where the potential of a user to bring new
users decreases through time. This suggests a possible ``user
expiration date'' after which a user of a OSN cannot be expected to
bring new users.

\subsection{Resilience and decline of Friendster}

We combined the sequence of user ids with the k-core decomposition of
\texttt{Friendster} to study how its resilience changed over time. In
particular, we explored the relation between the coreness of users and
the time when they joined the community. We divided users along the
median of the distribution of coreness values, $\bar{k_s}=6$. This
way, for each period of time, there is a set of users in the lower
half of the distribution ($k_s < \bar{k_s}$), which are nodes at risk
of leaving the OSN. We measure the resilience of these time-dependent
parts of network as the ratio between users at risk of leaving, and
the total amount of users in the slice.

\begin{figure}[h]
  \centering
    \includegraphics[width=0.8\textwidth]{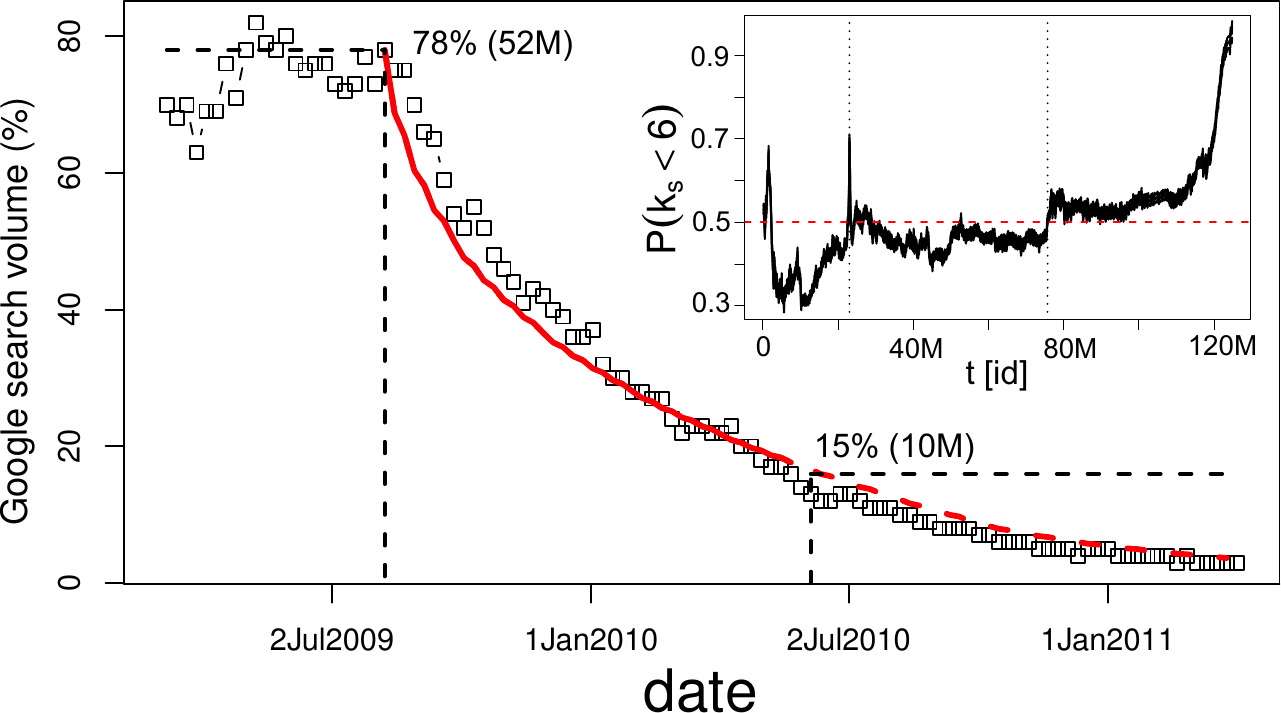}
    \caption{Weekly Google search trend volume for
      \texttt{Friendster}. The red line shows the estimation of the
      remaning users in a process of unraveling. Inset: time series of
      fraction of nodes with $k_s<6$.}
  \label{fig:fit}
\end{figure}

We created slices of $100000$ user ids, calculating a point sample
estimate of $P(k_s < \bar{k_s})$.  Inset of Figure
\ref{fig:fit} shows the time evolution of this ratio, with a dark
area showing 99\% confidence intervals.  First, we notice that the
skewness of $k_s$ does not affect our statistic, as the confidence
intervals are sufficiently concentrated around the point
estimates. Second, whe can identify certain time periods when the new
users of \texttt{Friendster} only connected to its fringe, having
larger ratios of nodes at risk. The first moment with a peak is at the
very beginning, to drop to ratios around 0.3 soon after. This shows
that the set of very early users did not fully exploit the social
network, and it took a bit of time for the OSN to become more
resilient. The second peak is shortly after having 22 million users,
which coincides with the decay of popularity of \texttt{Friendster} in
the US. Finally, the ratio of users at risk went above 0.5 before the
community had 80 million accounts, showing a lack of cohesion as its
shutdown approaches, as new users do not manage to connect to the
rest. 

To conclude our analysis, we explored how the spread of departures
captured in the k-core decomposition (see Section \ref{sec:model}) can
describe the collapse of \texttt{Friendster} as an OSN. As we do not
have access to the precise amount of active users of
\texttt{Friendster}, we proxy its value through the \texttt{Google}
search volume of \emph{www.friendster.com}. The inset of Figure
\ref{fig:fit} shows the relative weekly search volume from 2004, where
the increase of popularity of \texttt{Friendster} is evident. At some
point in 2009, \texttt{Friendster} introduced changes in its user
interface, coinciding with some technical problems, and the rise of
popularity of \texttt{Facebook}
\footnote{www.time.com/time/business/article/0,8599,1707760,00.html}. This
led to the fast decrease of active users in the community, ending on
its discontinuation in 2011.

We scale the search volumes fixing 100\% as the total amount of users
with coreness above $0$, 68 million.  At the point when the collapse
of \texttt{Friendster} started, the search volume indicates a
popularity of 78\% of its maximum.  We take this point to start the
simulation of a user departure cascade, with an initial amount of 58
million active users, i.e.  users with coreness above 3.  The second
reference point we take is June 2010, when Friendster was reported to
have 10 million active users
\footnote{en.wikipedia.org/wiki/Friendster}, corresponding to 15\% of
the 68 million user reference explained above. The search volume on
that date is 14\%, showing the validity of the assumption that the
maximum amount of active users corresponds to those with coreness
above 0. Thus, these 10 million remaining users correspond to nodes
with $k_s>67$.

Given these two reference points, we can approximate the collapse of
the network through its ``unraveling'' per k-core. Our assumption is
that a critical coreness $K_t$ starts at $3$ and increases by $1$ at a
constant rate. Such $K_t$ is the result of an increasing
cost-to-benefit ratio, and thus all the nodes with $k_s<K_t$ would
leave the community. Then, for each timestep, the amount of remaining
users would correspond to the CCDF shown in Figure
\ref{fig:coreDeg}. In our analysis, $K$ increases at a rate of $6$
per month, i.e. from $3$ to $67$ between our two reference point.

The red line of Figure \ref{fig:fit} shows the remaining users under
this process, with dashed values after the second reference point of
June 2010. We can observe that this process approximates well the
decay of \texttt{Friendster} from the start of its decline, to its
total shutdown in 2011. The $R^2$ value for this fit is $0.972$,
leaving some slight underfit through 2009.  This fit show the match
between two approximations: on one side the search volume as an
estimation of the amount of active users, and on the other side the
amount of remaining users when the $c/b$ ratio increases constantly
through time.

\section{Discussion}

In this article, we have presented the first empirical analysis of
social resilience in OSN.  We approached this question using a
theoretical model that relates the environment of the OSN with the
cascades of user departures. We showed how a generalized version of
the k-core decomposition allows the empirical measurement of
resilience in OSN.

We provided an empirical study of social resilience across five
influential OSN, including successful ones like \texttt{Facebook} and
unsuccessful ones like \texttt{Friendster}.  We have shown that the
hypothesis of a power-law degree distribution cannot be accepted for
any of these communities, discarding the epidemic properties of
complex networks as a possible explanation for large-scale cascades of
user departures.  Our k-core analysis overcomes this limitation,
quantifying social resilience as a collective phenomenon using the
CCDF of node coreness.  We found that the topologies of two successful
sites, \texttt{Livejournal} and \texttt{Facebook}, are less resilient
than the unsuccessful \texttt{Friendster} and \texttt{Orkut}.  This
indicates that the environmental condition of an OSN play a major role
for its success.  Thus, we conclude that the topology of the social
network alone cannot explain the stories of success and failure of the
studied OSN, and it is necessary to focus future empirical analysis in
measuring these costs and benefits. Additionally, we found very high
maximum coreness numbers for most of the OSN we studied. The existence
of these superconnected cores indicates that information can be spread
efficiently through these OSN \cite{Kitsak2010}.

As a case study, we provided a detailed post hoc analysis of the
changes in \texttt{Friendster} through time.  We detect that the range
of connections towards future nodes is much lower than the expectation
from a random process.  Using the coreness of the nodes, we could
track the time dependence of the risk of leaving for new users. We
found shocks that indicate periods of lower resilience of the whole
community.  Finally, we apply all our findings to
\texttt{Friendster}'s collapse, fitting an approximated time series of
active users through the spread of user departures predicted by the
k-core decomposition.  We estimated the amount of active users through
search volumes, but other sources can provide more reliable data, like
\texttt{Alexa} ranks, or last login times if provided by the
site. Such datasets would allow further validations of the k-core
decomposition as a measure of social resilience.

Our analysis focused on the macroscopic resilience of OSN, but
additional research is necessary to complete our findings. Microscopic
data on user activity and churn can provide estimators for the
benefits and costs of each network, to further validate the work
presented here. Furthermore, the generalized k-core can be applied
when user decisions are more complex than just staying or leaving the
network, for example introducing heterogeneity of benefits or weights
in the social links.

Another open question is the role of directionality in the social
network, and how to measure resilience when asymmetric relations are
allowed. The benefits of users of these networks would be
multidimensional, representing both the reputation of a user and the
amount of information it receives from its neighborhood.  The work
presented here is theoretically limited to the study of monotonously
increasing, convex objective functions of benefit versus active
neighborhood. While empirical studies support this assumption
\cite{Backstrom2006,Wu2013}, it is possible to imagine a scenario
where information overload decreases the net benefit of users with
very large neighborhoods, creating nonlinearities where the
generalized k-core is not a stable solution. We leave this questions
open for further research, and the study of social resilience in other
types of online communities.

\section{Acknowledgements}
The authors would like to thank the Internet Archive for their work in
crawling and curating \texttt{Friendster}.

\bibliographystyle{abbrv}

\end{document}